\documentclass[prl]{revtex4}

\begin{document}

\title{Condensation phenomena in gravity\\}

\author{Giorgio Papini}

\altaffiliation[Electronic address:]{papini@uregina.ca}
\affiliation{Department of Physics and Prairie Particle Physics Institute, University of Regina, Regina, Sask. S4S 0A2, Canada}%

\begin{abstract}

Gravity can play a role in critical phenomena. Topological singularities
induce ground state degeneracy and
break the continuum symmetry of the vacuum. They also generate
momenta oscillations about an average momentum and a positive
gravitational susceptibility. Gravitational analogues of the laws of Curie and Bloch
have been found for a one-dimensional model.
The critical temperature for a change in phase from unbound to isolated
vortices has also been calculated in a $XY$-model.

\end{abstract}

\maketitle


%
%




%

\section{1. Introduction}
Einstein equations tie the geometry of space-time to the distribution
of matter through the energy-momentum tensor which describes matter in definite states, fluid, or solid, or dust.
Matter, however, is not always in a definite state, but undergoes phase transformations that
are of great interest in physics. At a certain temperature, for instance, some
conductors become superconductors, some fluids superfluids and some solids ferromagnetic.
These critical phenomena share common properties. They occur near a critical
temperature $T_{c}$, under fixed conditions different phases can coexist and are characterized
by an order parameter, which is zero above $T_{c}$ and has nonzero and (possibly)
different values for different equilibrium states for $T<T_{c}$.

The interaction of gravity with matter at or near a critical point is not known.
In view of the importance of phase transitions, it is interesting to learn how space-time and matter interact
in these conditions. Aspects of the question are addressed below.

When a continuous symmetry is broken spontaneously in a quantum mechanical system, the
system's ground state evolves to a lower equilibrium configuration. This process is
at the heart of condensation.

Consider a system $\phi(x)$ represented by the covariant Klein-Gordon equation. Similar considerations can be extended
to systems described by other wave equations, but the essential points can be made neglecting spin.
On applying the Lanczos-De Donder condition
\begin{equation}\label{L}
  \gamma_{\alpha\nu,}^{\,\,\,\,\,\,\,
  \nu}-\frac{1}{2}\gamma_{\sigma,\alpha}^\sigma = 0 \,,
  \end{equation}
the covariant Klein-Gordon equation becomes, to $\mathcal{O}(\gamma_{\mu\nu})$
\begin{equation}\label{KG}
\left(\nabla_{\mu}\nabla^{\mu}+m^2\right)\phi(x)\simeq\left[\eta_{\mu\nu}\partial^{\mu}\partial^{\nu}+m^2
+\gamma_{\mu\nu}\partial^{\mu}\partial^{\nu}
\right]\phi(x)=0\,.
\end{equation}
Units $\hbar=c=k_{B}=1$ are used and the notations are as in \cite{PAP5}.
In particular, partial derivatives with respect to a variable $x_{\mu}$ are interchangeably
indicated by $\partial_{\mu}$, or by a comma followed by $\mu$. At this stage, the introduction
of quantum field theory is not necessary and use
is made of the external field approximation \cite{PAP1,PAP0} that treats gravity as a classical
theory when it interacts with quantum particles.
This approximation can be applied successfully to all those problems
involving gravitational fields of weak to intermediate strength for which the full-fledged use of general relativity
is not required \cite{PASP,PAP5,PAP2,LAMB,PAP6,PAP9,PAP3,PAP4}. These include
theories in which acceleration has an upper limit \cite{CAI1,CAI2,BRA,MASH1,MASH2,TOLL,SCHW,PU}
and that resolve astrophysical and cosmological singularities \cite{ROV,BRU},
theories of asymptotically safe gravity \cite{CA} and space-time deformation \cite{CapLamb}.

The first order solution of (\ref{KG}) is
\begin{equation}\label{PHA}
\phi(x)=\left(1-i\Phi_{G}(x)\right)\phi_{0}(x)\,,
\end{equation}
where $\phi_{0}(x)$ is a plane wave solution of the free Klein-Gordon equation,
\begin{equation}\label{PHI}
\Phi_{G}(x)=-\frac{1}{2}\int_P^x
dz^{\lambda}\left(\gamma_{\alpha\lambda,\beta}(z)-\gamma_{\beta\lambda,\alpha}(z)\right)
\left(x^{\alpha}-z^{\alpha}\right)k^{\beta}
\end{equation}
\[+\frac{1}{2}\int_P^x dz^{\lambda}\gamma_{\alpha\lambda}(z)k^{\alpha}=\int_{P}^{z}dz^{\lambda}K_{\lambda}(z,x)\,,\]\,
$P$ is an arbitrary point that may be dropped and
\begin{equation}\label{K}
K_{\lambda}(z,x)=-\frac{1}{2}\left[\left(\gamma_{\alpha\lambda,\beta}(z)
-\gamma_{\beta\lambda,\alpha}(z)\right)\left(x^{\alpha}-z^{\alpha}\right)-\gamma_{\beta\lambda}(z)\right]k^{\beta}\,.
\end{equation}
Equations (\ref{PHA}) and (\ref{PHI}) represent a space-time transformation of the vacuum
that makes the ground state degenerate, breaks the continuous symmetry of the
system and leads to the phenomenon of condensation.

To any order $n$ of $\gamma_{\mu\nu}$, the solution of (\ref{KG}) can be
written as
\begin{equation}\label{PHIn}
\phi(x)=\Sigma_{n}\phi_{(n)}(x)=\Sigma_{n}e^{-i\hat{\Phi}_{G}}\phi_{(n-1)}\,.
\end{equation}
where $\hat{\Phi}_{G}$ is the operator obtained from (\ref{PHI}) by replacing $k^{\alpha}$
by $i \partial^{\alpha}$. In what follows, use is made of the approximation $ \mathcal{O(\gamma_{\mu\nu})}$
for simplicity.

Two-point functions like (\ref{PHI}) belong to the family of world-functions introduced into tensor calculus by Ruse \cite{RU} and Synge \cite{SY}
and are used in general relativity to study the curvature structure of space-time. Here they also introduce elements of the topology of
space-time, as can be seen by taking the derivatives of $\Phi_{G}$ with respect to $z$. One obtains
\begin{equation}\label{PD}
\frac{\partial \Phi_{G}(z)}{\partial z^{\sigma}}=-\frac{1}{2}\left[\left(\gamma_{\alpha\sigma,\beta}(z)-\gamma_{\beta\sigma,\alpha}(z)\right)
\left(x^{\alpha}-z^{\alpha}\right)-\gamma_{\beta\sigma}(z)\right]k^{\beta}\,,
\end{equation}
which coincides with (\ref{K}) and
\begin{equation}\label{PD2}
\frac{\partial^{2}\Phi_{G}(z)}{\partial z^{\tau}\partial z^{\sigma}}-\frac{\partial^{2}\Phi_{G}(z)}{\partial z^{\sigma}\partial z^{\tau}}=R_{\alpha\beta\sigma\tau}\left(x^{\alpha}-z^{\alpha}\right)k^{\beta}\equiv\left[\partial z_{\tau},\partial z_{\sigma}\right]\Phi_{G}(z)=
K_{\sigma,\tau}(z,x)-K_{\tau,\sigma}(z,x)=\tilde{F}_{\tau\sigma}(z)\,,
\end{equation}
where $R_{\alpha\beta\sigma\tau}$ is the linearized Riemann tensor \cite{PAP10,PAP11}.
It follows from (\ref{PD2}) that $\Phi_{G}$ is not single-valued and that, after a gauge transformation, $K_{\alpha}$  satisfies the equations
\begin{equation}\label{DIV}
\partial_{\alpha}K^{\alpha}=\frac{\partial^{2}\Phi_{G}}{\partial z_{\sigma}\partial z^{\sigma}}=0
\end{equation}
and
\begin{equation}\label{2der}
\partial^2 K_{\lambda}=-\frac{k^{\beta}}{2}\left[\left(\partial^2 (\gamma_{\alpha\lambda,\beta})-\partial^2(\gamma_{\beta\lambda,\alpha})\right)\left(x^\alpha -z^\alpha\right)+
\partial^2 \gamma_{\beta\lambda}\right]
\end{equation}
identically, while the equation
\begin{equation}\label{PD3}
\left[\partial z_{\mu},\partial z_{\nu}\right]\partial z_{\alpha}\Phi_{G}=-\left(\tilde{F}_{\mu\nu,\alpha}+\tilde{F}_{\alpha\mu,\nu}+\tilde{F}_{\mu\alpha,\nu}\right)=0\,,
\end{equation}
holds everywhere. Therefore, $K_{\alpha}$ is regular everywhere, but $\Phi_{G}$ is
singular.

Differentiating $\Phi_{G}$ with respect to $x$ yields the term $\Phi_{G,\mu}$ that,
added to $k_{\mu}$, gives the generalized momentum
\begin{equation}\label{mom}
P_{\mu}=k_{\mu}+\Phi_{G,\mu}=k_{\mu}+\frac{1}{2}\gamma_{\alpha\mu}k^{\alpha}
-\frac{1}{2}\int^{x}dz^{\lambda}\left(\gamma_{\mu\lambda,\beta}(z)-
\gamma_{\beta\lambda,\mu}(z)\right)k^{\beta}\,,
\end{equation}
and the relation
\begin{equation}\label{M}
\frac{1}{2}\gamma_{\mu\nu}k^{\mu}k^{\nu}=\left(P_{\nu}-k_{\nu}\right)k^{\nu}=k^{\mu}\Phi_{G,\mu}=\frac{d\Phi_{G}}{d\tilde{s}}\,,
\end{equation}
where $\tilde{s}$ is the affine parameter along the world-line of $m$. It therefore seems natural to identify $\Phi_{G,\mu}$ with
the order parameter and $d\Phi_{G}/d\tilde{s}$ with the condensation force.

The transformation (\ref{PHA}) that makes the ground state of the system space-time dependent, produces a breakdown
of the vacuum symmetry and generates quantized oscillations of the ground state of the system. These are the Nambu-Goldstone bosons
that arise as a consequence of the vacuum degeneracy.
By substituting (\ref{PHA}) into the energy function $\tilde{H}=g_{\mu\nu}(k^{\mu}+\nabla^{\mu}\Phi_{G}(x))(k^{\nu}+\nabla^{\nu}\Phi_{G})$ and
using (\ref{mom}), one gets, to $\mathcal{O}(\gamma_{\mu\nu})$,
\begin{equation}\label{EXC}
\tilde{H}=g_{\mu\nu}P^{\mu}P^{\nu}=m^{2}+2\gamma_{\mu\nu}k^{\mu}k^{\nu}\,,
\end{equation}
that is, the fluctuations $\Phi_{G}$ about the original symmetric state $\phi_{0}$ are the
Nambu-Goldstone bosons that are massless because $\partial^{2}\Phi_{G}/(\partial z_{\sigma}\partial z^{\sigma})=\partial_{\alpha}K^{\alpha}=0$.

Dropping the unnecessary term $m$, equation (\ref{EXC}) gives the energy function
\begin{equation}\label{HE}
H= -\frac{1}{m}\gamma_{\mu\nu}k^{\mu}k^{\nu}\,.
\end{equation}

This introduction is followed by the calculation of the effect of gravity on
some physical quantities in a
one-dimensional model and by a calculation of the transition temperature in
a two-dimensional XY-model. This is the lowest dimension in which a finite calculation
can be carried out. A summary and conclusions are given in the last section.

\section{2. The one-dimensional model}
A lattice gas model can be used to calculate the alignment per particle and correlation
length that follow from gravity induced condensation. Some results are being repeated
here for the sake of completeness \cite{PAP12}.

The properties of a many-particle system satisfying (\ref{PHA}) follow from
$H$ which strongly resembles the energy function of the Ising model. A difference is represented here by the vectors $k_{\alpha}$
(or $P_{\alpha}$) that replace in (\ref{EXC}) the Ising spin variables
$\sigma_{i}$ which are numbers that can take the values $\pm 1$. It is however known that a
lattice gas model \cite{BAX}, equivalent to the Ising model, can be set up in which
the particles are restricted to lie only on the $N$ sites of a fine lattice, instead of being allowed to occupy any position in space-time.
Then one can associate to each site $i$ a variable $s_{i}=(1+\sigma_{i})/2$ which takes the value $1$ if the site is occupied by a vector $k_{\alpha}$ and
the value $0$ if it is empty. Any distribution of the particles can be indicated by the set of their site occupation numbers
${s_{1},. ...s_{N}}$.  By replacing
$k^{\mu}k^{\nu}$ in (\ref{EXC}) with their average $k^2 \eta^{\mu\nu}/4$ over the angles and restricting the interaction to couples of nearest sites,
one obtains
\begin{equation}\label{HH}
H=-\frac{m}{4} \gamma\left(\sum_{k=1}^{N}s_{k}s_{k+1}\right)\,,
\end{equation}
where $\gamma\equiv \gamma_{\mu\nu}\eta^{\mu\nu}$.
By imposing periodic conditions $s_{N+1}=s_{1}$ along the hypercylinder with axis parallel to the time-axis,
the partition function becomes
\begin{equation}\label{Z}
Z=\sum_{s_{1}}...\sum_{s_{N}}\exp\left(\beta \epsilon\sum_{k=1}^{N}s_{k}s_{k+1}\right)\,,
\end{equation}
where $\beta\equiv1/T$ and $\varepsilon\equiv m \gamma/4$ contains the gravitational contribution due to $\gamma_{\mu\nu}$.
This one-dimensional Ising model
can be solved exactly \cite{BAX}.
Equation (\ref{Z}) can be rewritten as $Z=\sum_{s_{1}}<s_{1}|\tilde{M}^{N}|s_{1}>=Tr(\tilde{M}^{N})=\lambda_{+}^{N}+\lambda_{-}^{N}$ and
the eigenvalues of $\tilde{M}$ are $\lambda_{+}=2 \cosh(\beta\varepsilon)$ and $\lambda_{-}=2\sinh(\beta\varepsilon)$. As $N\rightarrow \infty$,
$\lambda_{+}$ makes a larger contribution than $\lambda_{-}$, $N^{-1}\ln Z\rightarrow \ln(\lambda_{+})$ and the Helmholtz free energy per site is
$F/N=-(N \beta)^{-1}\ln Z\rightarrow -\beta^{-1}\ln(\lambda_{+})$. The alignment per particle for large values of $N$ is
\begin{equation}\label{AA}
\Gamma \equiv-\frac{1}{N}\frac{\partial F}{\partial\varepsilon}\sim -\frac{1}{\beta}\frac{d\ln \lambda_{+}}{d\varepsilon}=\tanh(\beta\varepsilon)\,,
\end{equation}
which yields the gravitational correction due to $\varepsilon$. It also follows from (\ref{AA}) that there is no spontaneous momentum alignment ($\Gamma=0$ when $\varepsilon=0$)
and that complete alignment $\Gamma=1$ is possible only for $T\rightarrow 0$.
In fact $F\rightarrow -N\varepsilon $ in the limit $T\rightarrow 0$ for completely aligned momenta and one can say that there is a phase transition
at $T=0$, but none for $T>0$.
It follows from (\ref{AA}) that the value of $\Gamma$ depends on $\gamma$.
It also follows that there is no alignment
($\Gamma =0$) for $T\rightarrow\infty$ (for any $\gamma$ and $m$), or for $\gamma =0$ (no gravity and any $T$).
According to (\ref{AA}), complete alignment $\Gamma =1$ can be achieved only at $T=0$
which, as shown below, is not, however, a critical temperature in the model.

The correlation length per unit of lattice spacing is
\begin{equation}\label{CSI}
\xi\sim\frac{1}{2}\exp(2\beta\varepsilon)\,,
\end{equation}
which gives $\xi=\infty$ at $T=0$ and $\xi=1/2$ at $T=\infty$ where thermal agitation can effectively disjoin neighbouring sites.

One can also define a gravitational susceptibility as $d\Gamma/d\gamma$. One finds
\begin{equation}\label{SUS}
\frac{d\Gamma}{d\gamma}=\frac{m\beta}{4\cosh^2 (\beta\epsilon)}\geq 0 \,
\end{equation}
always.
It follows that when $\beta\epsilon \ll 1$,
$\tanh (\beta\epsilon)\approx \beta\epsilon = \gamma m/4T$, and
\begin{equation}\label{SUS1}
\frac{d\Gamma}{d\gamma} \simeq \frac{m}{4T}= 2.9\times 10^{12}\frac{m(GeV)}{T(K)} \,,
\end{equation}
which is reminiscent of Curie's law.

A slightly better understanding of the problem can be obtained by calculating the value of $T$ for which
$1/\cosh^{2}(\beta\varepsilon)$ has a maximum. The derivative of $d\Gamma/d\gamma$ with
respect to $T$ vanishes when $2y \tanh(y)=1$ which gives $y=m \gamma/4T=0.77$. The susceptibility has a
maximum at
\begin{equation}\label{TM}
T_{m}(K)=\frac{m(GeV)\gamma}{2.65\cdot 10^{-13}} \,,
\end{equation}
and the corresponding value of the susceptibility is
\begin{equation}\label{SM}
\left(\frac{d\Gamma}{d\gamma}\right)_{m}=\frac{m}{4T_{m}\cosh^{2}(\frac{m\gamma}{4T_{m}})}=\frac{0.45}{\gamma}\,,
\end{equation}
which is independent of $m$. Table I lists the values of $(d\Gamma/d\gamma)_{m}$ for
some relevant astrophysical objects. In $T_{m}$, the nucleon mass $m\sim 0.9 GeV$ has been used
for simplicity. In general, even a small value of $\gamma$
is sufficient to saturate the alignment of momenta over a range of temperatures which is narrow
because of the sharpness of $d\Gamma/d\gamma$. At $T=T_{m}$, the correlation length is only $\xi\simeq 2.33$ which is rather small.
This is expected because the values $T=0$ and $T \simeq T_{m}$ correspond respectively to states of low and high
thermal agitation in which the system changes from high to small correlation.
It is therefore necessary to consider the value of $\xi$ at a particular temperature.
In all cases where $\beta\varepsilon\ll 1$, one finds $T\gg 1$ and $(d\Gamma/d\gamma) \ll 1/\gamma$. This is the high $T$
case. If $\beta\varepsilon \gg 1$, then $ T \ll (1/4) \gamma m $ and $0 \leq (d\Gamma/d\gamma) < 1.2 \cdot 10^{12}m(GeV)/T(K)$.
For Earth, $T \sim 300K$ yields
$d\Gamma/d\gamma \sim 1.6\cdot10^{7}$ and $\xi\sim 1.8\cdot 10^{7}$. It should be possible
to observe the condensation phenomenon in these conditions.

 \begin{table}
 \caption{ Maximum values of $ d\Gamma/d\gamma$ for some astrophysical objects.}
 \begin{tabular}{|c||c|c|c|}
 \,\, & $ \gamma $ & $T_{m} $ & $(d\Gamma/d\gamma)_{m} $\\
 \tableline 
 Earth & $ 10^{-9} $ & $ 3391 $ & $ 4.5\cdot10^{8} $
 \\
Sun & $ 2\cdot 10^{-6} $ & $ 7.2\cdot 10^{6} $ & $2.3\cdot 10^{5} $\\
Neutron star & $ 0.26 $ & $ 8.8\cdot10^{11} $ & $1.73 $\\
White dwarf & $ 10^{-3} $ & $ 3.4\cdot 10^{9} $ & $ 450 $\\
\tableline
 \end{tabular}
 \end{table}

It also follows from Table I that condensation effects can be large even though $\gamma$ is small, provided $\xi$ maintains 
a reasonable value. This is not normally observed for gravitational
phenomena,

The oscillations of $P_{\mu}$ are similar to spin waves and obey the
dispersion relation \cite{KIT,PAP12}
\begin{equation}\label{DR}
\omega =\frac{ms \gamma}{2}\left(1-\cos \ell a\right)\,,
\end{equation}
where $s$ is the spin magnitude at a site, $\ell=|\vec{\ell}|$ the spin-wave momentum and $a$ the lattice spacing.
Upon quantization, spin waves give rise to quasi particles that,
by analogy with magnons, shall be called "gravons".
For oscillations of small amplitude and on using (\ref{DR}), one obtains $ \omega\simeq ms\gamma (\ell a)^2 /2 =\ell^2 /2m^{*} $,
where $m^{*}$ is taken as the gravon's effective mass. If the lattice subdivision is very fine so
that $a\sim m^{-1} $, then $m^{*}=m/s\gamma $ can become large for small $\gamma $
and the oscillation frequency of these waves is very low.
For sufficiently small changes of $\gamma $, the energy of a mode of energy $\omega_{\ell}=\ell^2 /2m^{*}$ and
$n_{\ell}$ gravons is that of a harmonic oscillator and the gravon distribution is that of Planck.
When $\omega_{\ell}\ll T$, the number of gravons per unit volume is
\begin{equation}\label{ST}
\sum_{\ell}n_{\ell}\simeq 8(0.0587)\left(\frac{T}{ms\gamma a^{2}}\right)^{3/2}\,,
\end{equation}
also reminiscent of Bloch $T^{3/2}$ law for magnetism. Curie's law and (\ref{ST}) suggest $T_{c}=0$ as a critical
temperature. As shown in the next section, this choice is, however, inappropriate.

Finally, the radiation spectrum of gravons produced
by a proton a distance $b$ from a star of mass $M$ can be calculated
from the power radiated in the process of $p\rightarrow p'+\tilde{\gamma}$
using the equation
\begin{equation}\label{W}
W=\frac{1}{8(2\pi)^2}\int
\delta^{4}(P-p'-\ell)\frac{|M|^2}{Pp'_0}d^3p' d^3\ell\,,
\end{equation}
where
\begin{equation}\label{M1sq}
\Sigma |M_{p\rightarrow p'\tilde{\gamma}}|^2
=e^2\left[-4(p'^{\alpha}\Phi_{G,\alpha})+8
(p^{\alpha}\Phi_{G,\alpha})\right]\,,
\end{equation}
and $\tilde{\gamma}$ represents the gravon. The process is similar
to that of magnon production by neutron scattering by a magnetic structure.
The quantum mechanical power spectrum of $\tilde{\gamma}$ is
\begin{equation}\label{PS}
\frac{dW}{d\ell}\simeq \frac{e^{2}\ell}{\pi}\left(\frac{GM}{b}\right)\,,
\end{equation}
where $p=|\vec{p}|> m_{p}$ is the momentum of the incoming proton
and $pGM/b <\ell\leq p$ to satisfy the requirement $|\cos \vartheta|\leq1$,
where $\vartheta$ is the emission angle . Results (\ref{DR}) and (\ref{PS}) agree for $\ell a\ll 1$.

If $\omega_{\ell}\ll T$, the total number of gravons can be written in the form
\begin{equation}\label{NT}
\tilde{N}\simeq \frac{4T}{ms \gamma a}\sim \frac{4TL}{ms\gamma a^{2}\tilde{n}}\,,
\end{equation}
where $\tilde{n}=N/V$ and $L$ is the typical size of the system.
From $\tilde{N}/N \leq 1$, one gets $T\leq ms\gamma \tilde{n} a^{2}/4L$ which can be satisfied
for sufficiently large $L$.

\section{3. A two-dimensional model}

Not all ordered phases can exist because the number of space dimensions
plays a role in phase changes. Changes to ordered phases can survive
only if they are stable against long wavelength fluctuations. Consider, for instance, a system of
particles that is invariant under translations in a  space $V$ of $d$ dimensions. Representing the deviation
of the particles from the equilibrium position by $q(x)=(1/V)\sum_{k,r}e^{ik\cdot x}q_{i}(k)$, where
$i$ indicates the normal modes and $k$ extends up to some value, the normal modes energy is $T=\omega_{i}^{2}(k)|q_{i}(k)|^{2}$
by the theorem of equipartition of energy. Then for $V\rightarrow\infty$,
$<q(x)^{2}>=\Sigma_{i}\int d^{d}k 2T/\omega_{i}^{2}$.
If the continuous symmetry has been broken spontaneously, the excitations
whose frequencies $\omega_{i}(k)\sim k$, vanish as $k\rightarrow 0$ and
give the low frequency limit $<q^{2}>\sim\Sigma_{i}\int dk k^{d-1}/k^2$ which diverges for $d\leq 2$.
Hence the lowest critical dimension is $d=2$, below which order is destroyed by long wavelength fluctuations.

Consider, therefore, the case d=2.
The quickest way to obtain the relevant expressions for a two-dimensional model
is to restart from (\ref{HH}) and replace the vectors
$s_{i}$ with classical vectors constrained to lie in the $s_{x}s_{y}$ - plane. Then $s_{i}=(s_{ix},s_{iy})$ and
$s_{ix}^2+s_{iy}^2 =1$ and
\begin{equation}\label{HH1}
H=-\frac{m\gamma s^2}{4}\Sigma_{i,j}\cos(\theta_{i}-\theta_{j})
\end{equation}
if $i,j$ are neighbours, $H=0$ otherwise. If the neighbouring angles are close in value, then, neglecting
an irrelevant constant,
$H\simeq -(m\gamma s^2 /8)\Sigma_{\vec{R}}\Sigma_{\vec{a}}\left[-\theta(\vec{r})+\theta(\vec{r}+\vec{a})\right]^2$,
where $\vec{r}+\vec{a}$ is the nearest neighbour of $\vec{r}$. Replacing the finite differences with derivatives one gets
\begin{equation}\label{HH2}
H= \frac{1}{8}m\gamma s^2 \int d^{2}r \vec{\nabla}\theta(\vec{r})\cdot \vec{\nabla}\theta(\vec{r}) \,.
\end{equation}
These excitations are vortices for which $\vec{\nabla}\theta = n \vec{a_{\theta}}/r$ and $n$ is an integer.
The energy of an isolated vortex is $E=(1/8) m\gamma s^2 \int_{a}^{L}dr/r =(1/8) m\gamma s^2 \ln (L/a)$, where $L$ is
the linear dimension of the vortex. The entropy associated with a single vortex is $S= \ln(L/a)^2$ and
the change in free energy due to the formation of a vortex is
\begin{equation}\label{G}
\Delta G = \left(\frac{1}{8}m\gamma s^2 n^2 -2T \right) \ln\frac{L}{a}\,,
\end{equation}
which is positive for
\begin{equation}\label{TC}
T<\frac{1}{8} m\gamma s^2 n^2 \equiv T_{c}\,.
\end{equation}
Isolated vortices do not therefore form for $T\leq T_{c}$.
At low temperatures the state of the system consists of an equilibrium density of bound vortices.
At $T> T_{c}$ the vortices become unbound and the condensed phase is destroyed.

The condition $\tilde{N}/N\leq 1$ gives in two dimensions
\begin{equation}\label{NT}
\tilde{N}\simeq \frac{4T}{\tilde{n}ms\gamma a^{2}}\ln(\frac{L}{a})\leq 1\,,
\end{equation}
which can be satisfied for $0\leq T\leq m\tilde{n}s\gamma a^{2}/4\ln(L/a)$ provided $L/a \geq 1$.

Given the important role of space dimensionality in critical phenomena,
one may wonder about the behaviour of $\tilde{N}$ in dimensions higher than two.
In three dimensions the condition $\tilde{N}/N \simeq 2T/(\pi \tilde{n}ms\gamma a^{3})\leq 1$
can be satisfied at all $T$ for sufficiently high values of $\tilde{n}$.

\section{4. Summary and conclusions}

Quantum particles in gravitational fields undergo phase changes that may affects their behaviour in physical processes.
The question is relevant, in principle, to astrophysics and cosmology where the presence, or absence of some reactions
can have important consequences.
Critical phenomena also offer the opportunity to investigate some aspects of the interaction
between physical systems and space-time at different levels,
from topology to changes in length scales.

Covariant wave equations have solutions (\ref{PHA}) and (\ref{PHIn}),
that are space-time dependent transformations of the vacuum. The resulting degeneracy of the ground states
produces Nambu-Goldstone excitations which break the rotational symmetry of the system. The
quasiparticles generated, or gravons, are oscillations that obey the dispersion relation (\ref{DR})
and have an effective mass.

In the one-dimensional Ising model considered, the order parameter is the
generalized momentum $\Phi_{G,\mu}$,
along which the particle momenta tend to align. The motion of the particles in the direction of $\Phi_{G,\mu}$ is along geodesics.
Off them, along the variable $z$, $\Phi_{G}$
is topologically singular and length scales change. The phase singularities
of $\Phi_{G}$ are quantized vortices \cite{BE} and the particle motion along the hypersurfaces $\tilde{F}_{\alpha\beta}\neq 0$
satisfies the equation of deviation. Phase singularities give rise to strings of silence in acoustics, lines of
magnetic flux in magnetism and to vortices in optics, in superfluids and superconductors.
The multivalued nature of $\Phi_{G}$ leads to the loss of a standard of length in the region of critical phenomena.
For space-time loops linking the regions of singularity one must have
\begin{equation}\label{F}
\oint_{\Gamma}dz^{\lambda}K_{\lambda}(z)=\oint_{\Gamma} dz^{\lambda}\frac{\partial\Phi_{G}}{\partial z_{\lambda}}=\int_{\Sigma}d\sigma^{\lambda\mu}\tilde{F}_{\lambda\mu}
=2n\pi\,,
\end{equation}
where $\Sigma$ is a surface bound by $\Gamma$. The loss of length scale represented by (\ref{F}) can be interpreted
geometrically by assuming that, after parallel transport along $\Gamma$, vectors not only change direction, but also length.
This extension of the notion of parallel transport forms the basis of the theories of Weyl \cite{WE}, Dirac \cite{DI} and Weyl-Dirac \cite{WOO}.
Outside the critical region the change in length still vanishes around paths $\Gamma$ that do not link any singularity.
The field $\tilde{F}_{\sigma\tau}$
vanishes on the line $x_{\sigma}-z_{\sigma}=0$ which is entirely occupied by $\phi_{0}$.
In addition, the choice $\partial^{2}\gamma_{\alpha\beta}\neq 0$ in (\ref{2der}), would give a Meissner-type
effect for $\tilde{F}_{\sigma\tau}$. The result is here analogous to that of
superconducting strings. It is also remarkable that a singularity in a quantum
mechanical wave function produces a field $\tilde{F}_{\sigma\tau}$ that is entirely classical.

The effect of gravity on some parameters that characterize the critical behaviour of a quantum
system of scalar particles can be calculated. For instance, the
gravitational susceptibility $d\Gamma/d\gamma$ is always positive and obeys a Curie-type law.
The susceptibility can be understood as a measure of the reaction of the system to
a gravitational field and is analogous to the magnetic susceptibility. There are,
of course, no gravitational dipoles in an ensemble of quantum particles because Einstein's
gravity is always attractive. There are only the particle momentum vectors
in the lattice gas model considered. Therefore $d\Gamma/d\gamma$ indicates only how
the momentum alignment per particle changes when the gravitational parameter $\gamma$ changes.
Since $d\Gamma/d\gamma$ is always positive, the response to changes in $\gamma$ always increases
and can be termed as "paragravitational". The increase is larger as $T\rightarrow 0$
because at lower temperatures thermal agitation subsides and correlation is preserved.

The number of quantized spin-waves, or gravons, per unit volume follows a $T^{3/2}$ law,
the emission cross-section is low
and the gravon spectrum depends linearly on the momentum $\ell$, both classically and
quantum mechanically, if $\ell a\ll1$.

In the one-dimensional case one cannot say that $T=0$ is the critical temperature of the model.
The lowest critical dimension is $2$ and the simplest two-dimensional system is the $XY$-model.
A critical temperature is in fact given by (\ref{TC}),
below which isolated vortices do not exist, but only bound vortices.
Symmetry breaking has a topological origin also in the $XY$-model, as conjectured \cite {COH}.

Gravity plays a role in critical phenomena. The singularity in the solution (\ref{PHA})
introduces vortices and induces condensation.


\end{document}